\documentclass[manuscript=article]{achemso}

\usepackage{chemformula} % Formula subscripts using \ch{}
\usepackage{hyperref}
\usepackage[T1]{fontenc} % Use modern font encodings
\usepackage{subcaption}
\usepackage{titlesec}

\author{Prabeen Kumar Pattnayak$^1$}
\author{Aloke Kumar$^1$}
\author{Gaurav Tomar$^1$}
\email{*gtom@iisc.ac.in}
\affiliation[Indian Institute of Science]
{$^1$Department of Mechanical Engineering, Indian Institute of Science, CV Raman road, Bengaluru, Karnataka-560012, India}

\title[Influence of Rotational Diffusion on Macromolecular Self-Assembly Kinetics]
  {Influence of Rotational Diffusion on Macromolecular Self-Assembly Kinetics}

\abbreviations{IR,NMR,UV}

\begin{document}



\begin{abstract}
Macromolecular self-assembly underlies a plethora of biological processes and provides a versatile route for fabricating functional soft materials. The kinetics of self-assembly in solution are inherently stochastic and are fundamentally governed by the interplay of translational and rotational diffusion of the constituent macromolecules. While most computational studies model macromolecules as patchy spherical colloids, thereby neglecting the influence of polymer architecture and internal conformational dynamics, the role of these factors in macromolecular self-assembly kinetics remains poorly understood. Here, we investigate the self-assembly of two patchy macromolecules with different architectures, namely linear chains and star polymers with four and seven arms. The hydrodynamic radii of the macromolecules are chosen to be nearly identical, thereby matching their translational diffusion coefficients and thus isolating the influence of rotational diffusion on the self-assembly process. The binding probability of the patchy macromolecules is found to depend strongly on their internal architecture. Furthermore, reactive path density analysis reveals that self-assembly pathways are influenced by the rotational diffusion coefficient of the individual macromolecules. Overall, this study establishes a bridge between the equilibrium dynamics of macromolecules and their self-assembly kinetics, highlighting the importance of polymer internal architecture in the process of self-assembly.
\end{abstract}


\section{Introduction}
Macromolecules constitute a major class of soft matter systems, whose structural organization and dynamics are governed by the interplay between intermolecular interactions and thermal fluctuations. Since the characteristic interaction energies are often comparable to the thermal energy, the self-assembly of macromolecules in solution is inherently stochastic, with thermal fluctuations continuously influencing macromolecular diffusion, encounters, orientations, and binding events. Macromolecular self-assembly is a fundamental process underlying numerous biological functions and advanced soft materials. Liposomes, which are formed from the self-assembly of lipid molecules, are used for drug delivery for cancer treatment\cite{zhou2021facile}. Rotaxanes, another self-assembled molecule, can be used as switches in molecular electronics\cite{yang2022roadmap,wu2022rotaxane,wang2026progress}. Virus capsids are a striking example of biological self-assembly, where capsomer proteins with specific interaction sites spontaneously organize into symmetric polyhedral structures\cite{hagan2006dynamic,bruinsma2003viral}.  Another biologically important example of self-assembly is the formation of protein complexes in cellular signal-transduction networks\cite{bray1995protein,law2025peptides}. Self-assembly provides a bottom-up route for material synthesis, wherein macromolecular building blocks spontaneously organize into ordered structures through specific intermolecular interactions\cite{hamley2023self,jawad2026nanomaterial}. The driving force for self-assembly is typically a cooperative combination of non-covalent interactions\cite{whitesides2002self,service2005far}, including van der Waals interactions, hydrogen bonding, $\pi$–$\pi$ interactions, and hydrophobic interactions. The relative contribution of these interactions depends on the specific system. For example, hydrophobic interactions predominantly drive the self-assembly of liposomes\cite{zhou2021facile}, virus capsids\cite{hagan2006dynamic,bruinsma2003viral}, and many protein complexes\cite{bray1995protein,law2025peptides}, whereas hydrogen bonding is central to the formation of many rotaxane-based assemblies\cite{yang2022roadmap,wu2022rotaxane}. In contrast, bottom-up material synthesis often exploits the synergistic action of multiple non-covalent interactions to direct the formation of ordered structures with tailored properties\cite{hamley2023self,jawad2026nanomaterial}. Self-assembly usually takes place in a fluid medium. Hydrodynamic interactions play a crucial role in the self-assembly of soft matter systems by mediating long-range forces between building blocks through the surrounding fluid\cite{grzybowski2000dynamic,fialkowski2006principles,soh2008dynamic}. These fluid-mediated interactions couple translational and rotational motion, thereby governing the assembly kinetics, collective dynamics, and the emergence of ordered structures. Consequently, hydrodynamic interactions provide an effective means of controlling self-assembly pathways and dynamic reconfiguration in colloidal, interfacial, and other soft matter systems.

The dynamical pathway of macromolecular self-assembly represents the routes followed by the system during the assembly process and is strongly dependent on how far apart the macromolecules are, governed by the translational diffusion of individual macromolecules, and on how they are oriented with respect to each other, governed by their rotational diffusion. Most of the self-assembly studies use spherical colloids as a model system for macromolecules\cite{nag2024dissipative,hendley2023multistate,das2021variational,prestipino2017self,newton2015rotational}. It has been shown how rotational diffusion affects the dynamical pathway followed by the patchy colloids during self-assembly\cite{newton2015rotational}. The modeling of macromolecular self-assembly by using patchy colloids provides an initial road map to study the effect of transport properties on self-assembly. However, the effect of internal dynamics of the macromolecules on self-assembly is not fully captured by modeling these as colloids. The internal architecture of the polymer is shown to play an important role in various dynamical processes\cite{su1991computer,korolkovas20185d,vahid2025collective,sarkar2018computational,zhang2019parallel,wang2025imperfect,upadhyay2025packing,wu2025conformation,pal2026molecular,rooks2026two,mahato2026anomalous}. It has been shown that macromolecules of various topologies, with equal effective hydrodynamic size, can exhibit significantly different rotational diffusion rates due to their distinct internal structures, expressed using shape anisotropy\cite{pattnayak2024diffusion,pattnayak2026reorientational}. Hence, in order to study the effect of internal dynamics of macromolecules on the dynamical pathway of self-assembly, it is essential to simulate macromolecular self-assembly by modeling the macromolecules explicitly. Self-assembly studies that model macromolecules of various architectures are scarce. 

\begin{figure}[t]
\centering
\includegraphics[scale=1]{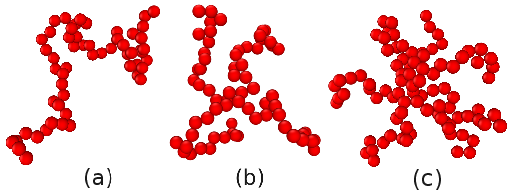}
\caption{Schematic representation of polymers under consideration: (a) linear, (b) star ($f=4$), and (c) star ($f=7$). All three polymers have approximately the same hydrodynamic radius ($R_h \sim 9.56 \sigma_p$). For visualization, OVITO\cite{stukowski2009visualization} has been used.}
\label{fig:schematic_polymers}
\end{figure}

In this work, we study macromolecular self-assembly by considering three cases, each with different polymer architectures: linear, star ($f=4$), and star ($f=7$), where $f$ is the functionality of the star polymer (see Figure \ref{fig:schematic_polymers}). The molecular weights of the macromolecules in the individual cases of self-assembly are selected so that their hydrodynamic radii are approximately constant, matching the translational diffusion rate, and thus isolating the effect of rotational diffusion on the dynamical pathways of macromolecular self-assembly. This study provides a connecting link between the internal dynamics of the macromolecules and their self-assembly kinetics, given that the rotational diffusion of macromolecules is highly sensitive to their internal structures\cite{pattnayak2024diffusion,pattnayak2026reorientational}. Such mechanistic insights are essential for predicting and optimizing self-assembly in biological systems and for the rational design of functional materials with improved assembly efficiency and yield.

\section{Numerical Formulation}

We have modeled the polymer chain as a series of identical monomer beads of mass $M$ connected by springs. The excluded volume interactions between the monomer beads are modeled by using the Weeks-Chandler-Andersen (WCA) potential\cite{weeks1971role} as follows,
\begin{equation}
U_{WCA}(r) = 
    \begin{cases}
  4\varepsilon \left[ \left( \frac{\sigma_p}{r} \right)^{12} - \left( \frac{\sigma_p}{r} \right)^6 \right] + \varepsilon & $r $\leq 2^{1/6}\sigma_p$ $ \\
  0 & \text{otherwise}
\end{cases}
\end{equation}
where $\sigma_p$ is the bead diameter, $r$ is the distance between two beads and $\varepsilon = k_BT$ ($k_B$ is Boltzmann's constant and $T$ is absolute temperature) is the interaction strength. The adjacent beads are connected to each other by springs with the finitely extensible nonlinear elastic (FENE) potential as follows,
\begin{equation}
    U_{FENE}(r) = 
\begin{cases}
  -\frac{1}{2} k r_0^2 \ln \left[ 1 - \left(\frac{r}{r_0}\right)^2 \right]   & $r $\leq r_0$ $ \\
  \infty & \text{otherwise}
\end{cases}
\end{equation}
where the standard Kremer-Grest parameters\cite{grest1986molecular,kremer1990dynamics} have been employed for the spring constant ($k = 30 k_BT / \sigma_p^2$) and maximum spring extension ($r_0 = 1.5 \sigma_p$).

We have used the hybrid molecular dynamics - multi-particle collision dynamics (MD-MPCD) method, which is a combination of molecular dynamics simulations for the macromolecules (solute) and the multi-particle collision dynamics method\cite{malevanets1999mesoscopic} for modeling solvent-mediated hydrodynamic interactions. In the MPCD simulations, the solvent is modeled explicitly as an ensemble of non-interacting point particles of finite mass ($m$). The motion of the solvent particles (or MPCD particles) are governed by alternating ballistic streaming and stochastic collisions. In the streaming step of time interval $\delta t$, the positions ($\textbf{r}_i$) of the solvent particles are updated as follows,
\begin{equation}
    \boldsymbol{r}_i( t + \delta t) = \boldsymbol{r}_i(t) + \delta t \boldsymbol{v}_i(t)
\end{equation}
where $\boldsymbol{v}_i$ is the velocity of solvent particle. During this streaming interval, the positions and velocities of the monomer beads are updated as per Newton's equations of motion, which are integrated numerically using the velocity-Verlet algorithm\cite{frenkel2023understanding,tuckerman1992reversible} with time step $\delta t_{MD}$. In the MPCD collision step, the simulation box is divided into identical cubic cells of size ($a$) equals to diameter of the monomer beads ($a = \sigma_p$), and the velocities ($\boldsymbol{v}_i$) of the particles within the cells are updated stochastically using the momentum-conserving Andersen Thermostat (MPCD-AT)\cite{allahyarov2002mesoscopic}, as follows,
\begin{equation}
    \boldsymbol{v}_i(t + \delta t) = \textbf{v}_{cm}(t) + \boldsymbol{v}_i^{ran} - \Delta \textbf{v}_{cm}^{ran}
\end{equation}
where $\textbf{v}_{cm}$ is the center-of-mass velocity of the collision cell, $\boldsymbol{v}_i^{ran}$ is a random velocity, the three components of which are selected randomly from a Gaussian distribution with variance $k_BT/m$ for the solvent particles and $k_BT/M$ for the monomer beads. $\Delta \textbf{v}_{cm}^{ran}$ is the change in center-of-mass velocity of the collision cell due to addition of $\boldsymbol{v}_i^{ran}$, and subtracting it ensures conservation of linear momentum. This kind of modeling the solvent-monomer interaction by including the monomer beads in the stochastic collision of MPCD is often used in recent studies\cite{hegde2011conformation,jiang2013accurate,nikoubashman2017equilibrium,chen2017effect,chen2018coupling,chen2021nanoparticle} due to its benefit of avoiding spurious depletion forces\cite{padding2006hydrodynamic}. The collision cells are shifted before every collision step by a vector with the three components chosen randomly from $[-a/2,a/2]$ for ensuring Galilean invariance\cite{ihle2001stochastic}. We have carried out the simulations using the MPCD-AT routines in LAMMPS\cite{plimpton1995fast,LAMMPS}(Chen et al.\cite{chen2018coupling,2017GitHub}).

The dynamic properties of the solvent are regulated by the average density of the MPCD particles ($\rho_s = 5 m / \sigma_p^3$), and the MPCD collision time step ($\delta t = 0.09\tau$), where $\tau$ is the intrinsic unit of time equals $\sqrt{m\sigma_p^2/k_BT}$. The mass of a monomer bead ($M$) is selected as 5$m$ to match the average solvent mass in the MPCD collision cell. The resulting dynamic viscosity ($\eta$) and the corresponding Schmidt number ($Sc$) of the present MPCD solvent are $4 \tau k_BT / \sigma_p^3$ and 12, respectively. The MD time step ($\delta t_{MD}$) is chosen to be $0.002\tau$. The measured parameters are expressed using the implicitly reduced units: energy scale $k_B T$, length scale $\sigma_p$, and mass scale $m$.

The translational and rotational diffusion coefficients are obtained by simulating the Brownian motion of a single polymer chain in the MPCD solvent in a cubic periodic simulation box. The size of the box is selected to be $54 \sigma_p$. The equilibration simulation run is performed for $2\times10^7$ MD time steps. The results are time averaged over $5\times10^8$ MD time steps and ensemble-averaged over five system replicas, each with a distinct set of random velocities at the beginning of the simulation and in the stochastic collision, both taken from the Maxwell-Boltzmann distribution. The selected polymers are linear and star polymers with functionalities of 4 and 7, respectively.

\begin{figure}
\centering
\includegraphics[scale=0.95]{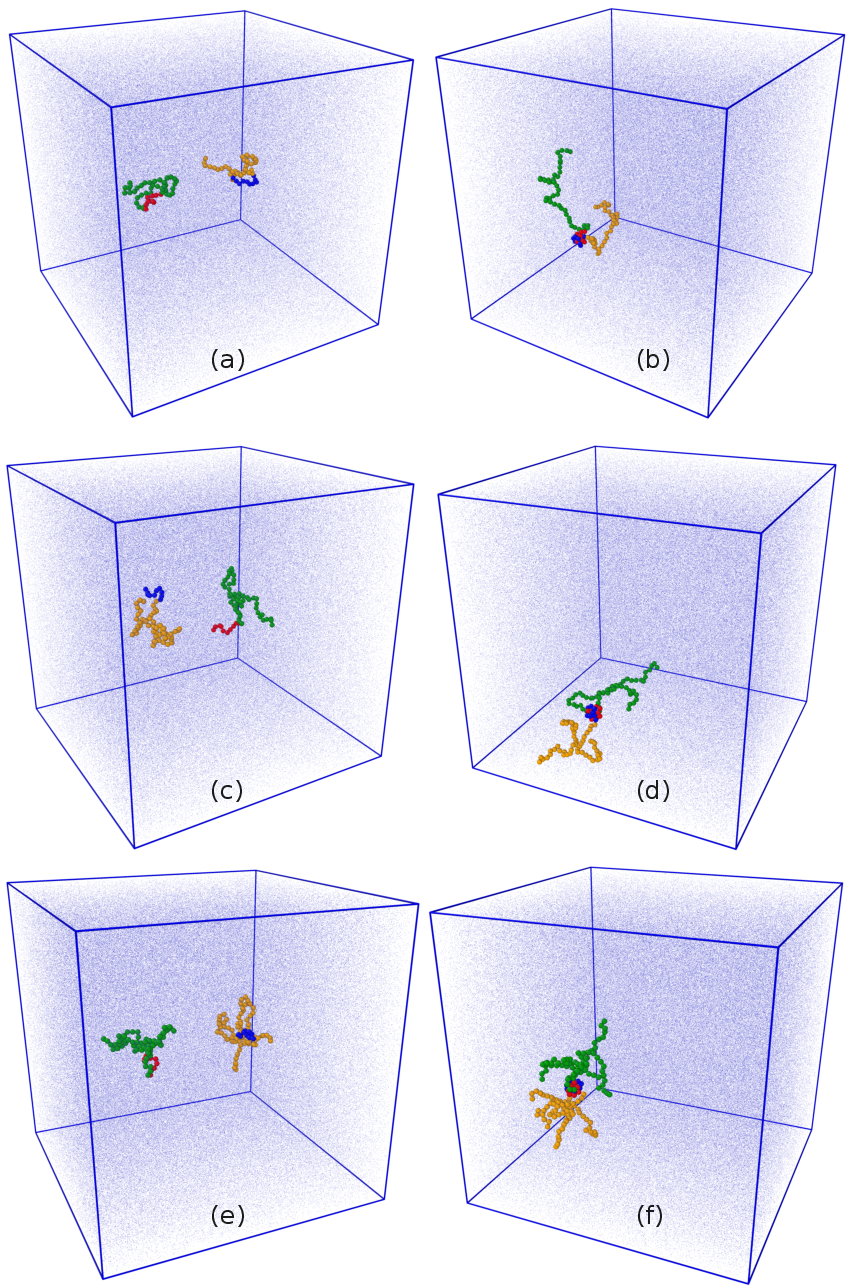}
\caption{MPCD simulation snapshots of polymer chains at initial configuration and after successful binding: linear chain ((a) and (b)), star polymer with functionality four ((c) and (d)), and star polymer with functionality seven ((e) and (f)). For visualization, OVITO\cite{stukowski2009visualization} has been used.}
\label{fig:selfassembly}
\end{figure}

The self-assembly simulations are performed by modeling a certain portion of the polymer as an attractive patch. The interactions between the inter-polymer attractive patches are modeled by introducing the 12–6 Lennard-Jones potential among the monomers of two different patches of the two polymers as follows,

\begin{equation} \label{eqn:LJ126}
U_{LJ}(r) = 
    \begin{cases}
  4\varepsilon \left[ \left( \frac{\sigma_p}{r} \right)^{12} - \left( \frac{\sigma_p}{r} \right)^6 \right] & $r $\leq 40 \sigma_p$ $ \\
  0 & \text{otherwise}
\end{cases}
\end{equation}
where $\varepsilon = 30 k_BT$ is the strength of interaction. The strength of the interaction potential is chosen to be significantly higher than the thermal energy of the solvent, with a large cutoff distance to enable the self-assembly of diffusing macromolecules in the finite simulation time\cite{newton2015rotational}. The patch portion of each polymer, across all three architectures, consists of 10 monomers. Two identical equilibrated polymer chains are placed in the MPCD solvent at a distance of $20 \sigma_p$ apart, which is approximately 4 times the radius of gyration of the individual polymers, in a cubic simulation box of size $54 \sigma_p$ and are allowed to evolve with time using Newton’s equations of motion for $2 \times 10^7$ MD time steps. Periodic boundary conditions are implemented in all directions. Different ensemble runs are carried out, each with a unique set of random velocities at the beginning of the simulation and during
stochastic collision of MPCD, both taken from the Maxwell-Boltzmann distribution. The simulation snapshots of some of the successful runs are shown in Figure \ref{fig:selfassembly}.

\newpage

\section{Results and Discussion}

The hydrodynamic radius ($R_h$) of a macromolecule, defined by preaveraging of the pairwise hydrodynamic interactions in the Zimm theory framework, is calculated as follows\cite{dunweg1991microscopic,dunweg1993molecular,dunweg1998molecular},
\begin{equation} \label{eqn:rh}
   \langle \frac{1}{R_h} \rangle = \frac{1}{N^2} \sum_{i \ne j} \langle \frac{1}{r_{ij}} \rangle
\end{equation}
where $r_{ij}$ is the distance between the monomers of a macromolecule and $N$ is its molecular weight. The molecular weights of the three types of polymer chains under consideration are selected such that the resulting hydrodynamic radius is approximately the same ($R_h \sim 9.56 \sigma_p$), as shown in the fourth column of Table \ref{tbl:ratioSE}. Another measure of the average size of the macromolecule is its radius of gyration ($R_g$), which is equal to the trace of the gyration tensor of the macromolecule\cite{theodorou1985shape,khabaz2014effect}. The radius of gyration measures the spatial extent of the polymer chain from its center of mass. The computed values of $R_g$ for the three types of macromolecules under consideration are listed in the third column of Table \ref{tbl:ratioSE}. The linear polymer has the highest value of $R_g$ followed by the star polymers with $f=4$ and $f=7$, respectively. This decreasing trend of $R_g$ with the increase in functionality at constant $R_h$ of the star polymers is in good agreement with the fact that the ratio, $R_g / R_h$ of star polymers, reduces with an increase in $f$\cite{burchard1980static,huber1984dynamic,singh2014hydrodynamic,pattnayak2024diffusion}. The distinct internal structure of a macromolecule leads to its peculiar shape, as expressed by relative shape anisotropy ($\kappa^2$), a dimensionless quantity, and is calculated as follows\cite{theodorou1985shape,khabaz2014effect},
\begin{equation} \label{eq:shape}
    \kappa^2 = 1 - 3 \frac{\lambda_1 \lambda_2 + \lambda_2 \lambda_3 + \lambda_3 \lambda_1}{( \lambda_1 + \lambda_2 + \lambda_3 )^2}
\end{equation}
where $\lambda_1$, $\lambda_2$, and $\lambda_3$ are the eigenvalues of the gyration tensor of the macromolecule. $\kappa^2$ ranges from 0, which represents a sphere or any of the five Platonic solids, to 1, representing a straight rod. The computed values of $\kappa^2$ of the three types of macromolecules under consideration are summarized in the sixth column of Table \ref{tbl:ratioSE}. The linear chain has the highest value of relative shape anisotropy, whilst the star polymer with $f=7$ has the lowest value.

\begin{table}[t]
  \centering
  \caption{Molecular weight ($N$), radius of gyration ($\bar{R}_g = \langle R_g^2 \rangle^{1/2}$), hydrodynamic radius ($\bar{R}_h = \langle 1 / R_h \rangle^{-1} $), translational diffusion coefficient ($D_t$), relative shape anisotropy ($ \kappa^2 $), rotational diffusion coefficient ($D_r$), and rotational diffusion ratio ($f_{SE}$).}
  \label{tbl:ratioSE}
  \begin{tabular}{lccccccc}
    Type of chain & $N$ & $\bar{R}_g/ \sigma_p$ & $\bar{R}_h/\sigma_p$ & $D_t/(\sigma_p^2/\tau)$ & 
    $\langle \kappa^2 \rangle$ &
    $D_r \tau$ &  $f_{SE}$ \\
    \hline
    Linear & 56 & $5.08 \pm 0.93$ & $9.58 \pm 1.77$ & $3.2 \times 10^{-3}$ & 0.46 & $1.4 \times 10^{-4}$ &  1.16\\
    Star ($f = 4$) & 68 & $4.46 \pm 0.43$ & $9.47 \pm 1.89$ & $2.9 \times 10^{-3}$ & 0.21 & $4.28 \times 10^{-4}$ &  2.1 \\
    Star ($f = 7$) & 91 & $4.2 \pm 0.24$ & $9.66 \pm 2.02$ & $2.5 \times 10^{-3}$ & 0.09 & $8.05 \times 10^{-4}$ & 3.17 \\
    \hline
  \end{tabular}
\end{table}

\subsection{Translational and rotational diffusion}
\begin{figure}
\centering
\includegraphics[scale=1]{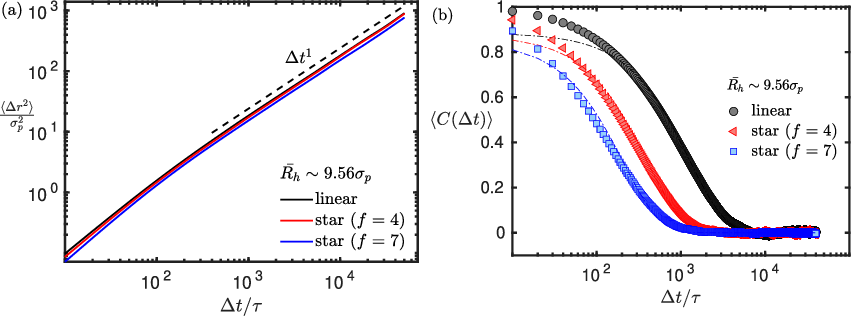}
\caption{(a) Variation of center-of-mass mean square displacement, $\Delta r^2$, with lag time and (b) Variation of reorientational correlation function, $C(\Delta t)$ with lag-time for linear and star polymers with functionality four and seven, and having approximately the same hydrodynamic radius, $R_h \sim 9.56\sigma_p$.}
\label{fig:msd_rcf}
\end{figure}

The rate of translational diffusion of a macromolecule can be measured by variation of the mean squared displacement (MSD) of its centre-of-mass with lag-time, as shown in Figure \ref{fig:msd_rcf}(a)  for the three types of polymers under consideration. In MSD ($\Delta r^2$) vs. lag-time ($\Delta t$) plot, after the initial inertial regime ($\Delta t < 300 \tau$), the linear diffusive regime is reached, from which the translational diffusion coefficients ($D_t$) of the respective macromolecules are calculated by fitting the relation, $\Delta r^2 = 6 D_t \Delta t$ using the least-square-method and are summarized in the fifth column of Table \ref{tbl:ratioSE}. There is a negligible difference in the values of the translational diffusion coefficients of the three types of macromolecules under consideration since all of them have approximately the same values of hydrodynamic radius. This is in very good agreement with Zimm theory\cite{dunweg1998molecular,pattnayak2024diffusion}, which states that the translational diffusion coefficient of a macromolecule is inversely proportional to its hydrodynamic radius.

\begin{figure}[t]
\centering
\includegraphics[scale=1]{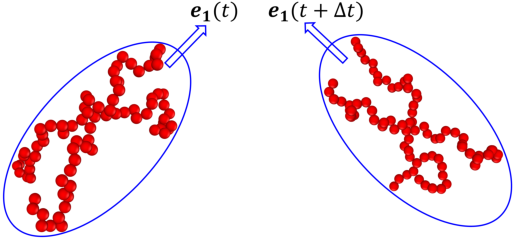}
\caption{Schematic illustration of reorientation of a polymer chain using an imaginary ellipsoid. The normalized eigenvector ($\boldsymbol{e_1}$), corresponding to the largest eigenvalue of the gyration tensor, is used to measure the reorientation rate of the polymer chain\cite{theodorou1985shape,pattnayak2024diffusion,pattnayak2026reorientational}.}
\label{fig:ellip}
\end{figure}

The rate of rotational diffusion of a macromolecule can be calculated by measuring the de-correlation of the principal axis of the gyration tensor with time\cite{theodorou1985shape,pattnayak2024diffusion,pattnayak2026reorientational} (see Figure \ref{fig:ellip}). The corresponding re-orientational correlation function (RCF)\cite{wong2009influence,pattnayak2024diffusion,pattnayak2026reorientational} is defined as, $C(\Delta t) = \langle P_2( \boldsymbol{e_1}(t).\boldsymbol{e_1}(t+\Delta t) ) \rangle$, where $\boldsymbol{e_1}$ is normalized vector representing the principal axis of the gyration tensor of the macromolecule, $P_2(x) = (3x^2 - 1)/2$ is the second-order Legendre polynomial, and the angle bracket represents the time and ensemble average over five system replicas. The decays of the respective RCFs of the three types of macromolecules with time are shown in Figure \ref{fig:msd_rcf}(b). The faster the decay of the RCF, the higher the rate of reorientation of the corresponding macromolecule. The RCF of the star polymer with $f=7$ decays the fastest, while that of the linear chain decays the slowest. The rotational diffusion coefficient ($D_r$) of a macromolecule can be calculated by fitting a single exponential to the decay of its RCF\cite{wong2009influence,pattnayak2024diffusion,pattnayak2026reorientational}, i.e., $C(\Delta t) = e^{-6 D_r \Delta t}$. The computed values of $D_r$ for the three types of polymers considered are listed in the second last column of Table \ref{tbl:ratioSE}. Despite having approximately the same hydrodynamic radius, the star polymers exhibit distinct rotational diffusion coefficients due to differences in their internal structure, as quantified by the relative shape anisotropy\cite{pattnayak2024diffusion}. A comparison of the translational and rotational diffusion coefficients of these polymers reveals a fundamental distinction between the two transport processes: where translational diffusion is largely insensitive to the internal architecture of the macromolecules at a fixed hydrodynamic radius, rotational diffusion is highly sensitive to the internal dynamics and conformational anisotropy of the macromolecule during Brownian motion\cite{pattnayak2024diffusion,pattnayak2026reorientational}.

\newpage

\subsection{Binding probability of self-assembly}
\begin{figure}[t]
  \includegraphics[scale=1]{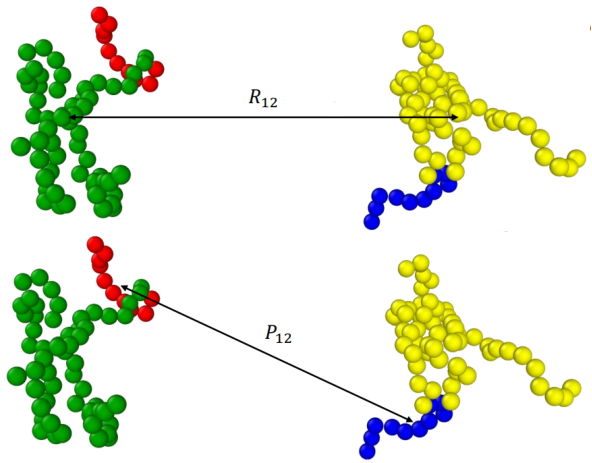}
\centering
  \caption{Schematic representation of two tracking variables: $R_{12}$, that is, distance between two centers-of-mass of the polymers and $P_{12}$, that is, distance between two centers-of-mass of the patches on the respective polymers.}
  \label{fig:r12p12}
\end{figure}

\begin{figure}[t]
\centering
\includegraphics[scale=1]{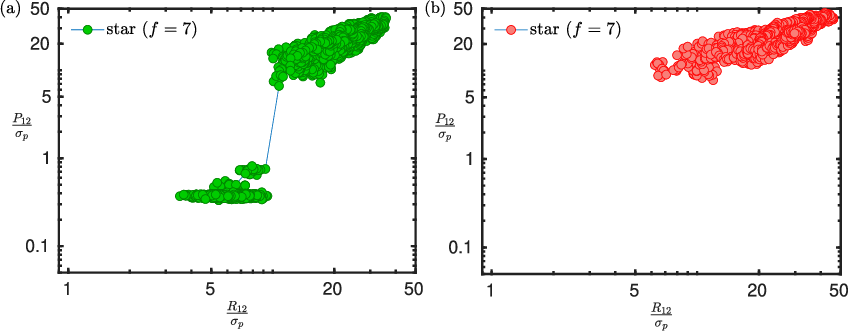}
\caption{Trajectory of self-assembly for two star polymers with functionality seven, (a) one of the successful runs and (b) one of the unsuccessful runs.}
\label{fig:star7_asemy}
\end{figure}

The self-assembly process can be monitored by calculating the distance between the centers of mass of the two polymers ($R_{12}$) and the distance between the centers of mass of the two patches ($P_{12}$) on the respective macromolecule units (see Figure \ref{fig:r12p12}). At the instance when the value of $P_{12}$ is less than the diameter of a monomer ($\sigma_p$), it is considered that binding has successfully taken place. Considering the fact that self-assembly is a stochastic process, the successful binding of the two macromolecule units may not occur in every simulation run within a fixed simulation time. The trajectories of successful and unsuccessful runs are shown in  Figure \ref{fig:star7_asemy} for star polymer (f=7). The number of successful runs for the three types of macromolecules under consideration are given in Table \ref{tbl:runselfassembly}. The probability of binding can be defined as $S/T$, where $S$ is the number of simulation runs where successful binding is observed within the prescribed simulation time and $T$ is the total number of independent simulation runs for a given polymer architecture. As shown in Table \ref{tbl:runselfassembly}, the binding probability is highest for the linear polymer and decreases for star polymers with functionalities $f=4$ and $f=7$, with the latter exhibiting the lowest value. Since all simulations are initiated from the same centers of mass distance ($R_{12} = 20\sigma_p$) in the periodic simulation box of the same size and the polymers possess nearly identical hydrodynamic radii, their center-of-mass translational diffusion coefficients, and consequently their encounter frequencies, are expected to be nearly identical. The observed differences in binding probability are therefore likely associated with differences in macromolecular architectures. Specifically, the linear polymer possesses a larger radius of gyration than the star polymers (third column of Table \ref{tbl:ratioSE}), resulting in a larger effective spatial extent and, consequently, a larger effective capture cross-section of its attractive patch during molecular encounters. As a result, the probability of forming a successful binding event decreases with decreasing $R_g$.
\begin{table}[t]
\centering
\caption{Probability of binding ($S/T$) in a given simulation time of $4 \times 10^4 \tau$: No. of total runs ($T$) and No. of runs with successful binding ($S$)}
\begin{tabular}{lccc}
Type of chain & T & S & $S/T$ \\
\hline
Linear chain & 111 & 96 & 0.87 \\
Star polymer ($f=4$) & 127 & 94 & 0.74 \\
Star polymer ($f=7$) & 155 & 91 & 0.59 \\
\hline
\end{tabular}
\label{tbl:runselfassembly}
\end{table}
These results demonstrate that, even for polymers with identical hydrodynamic radii and similar center-of-mass translational diffusion rates, subtle differences in internal architecture, reflected by the difference between $R_g$ and $R_h$, can significantly influence self-assembly kinetics. Such architecture-dependent effects are not captured by conventional rigid patchy-colloid models, which neglect the internal conformational dynamics of macromolecules. More broadly, these findings highlight that macromolecular self-assembly is governed not only by the translational diffusion of the building blocks but also by their internal conformational characteristics, emphasizing the importance of explicitly accounting for polymer architecture in coarse-grained models of self-assembly.

\subsection{Dynamical pathway of self-assembly}
Reactive path density provides a systematic way to characterize how a system proceeds from an initial state to a final assembled state, focusing on the dynamical pathways that lead to successful assembly\cite{newton2015rotational}. In the context of polymer self-assembly, where binding is often irreversible, and equilibrium assumptions may not hold, reactive path density offers a powerful framework to extract mechanistic insight from simulation trajectories.

\begin{figure}[t!]
  \includegraphics[scale=1]{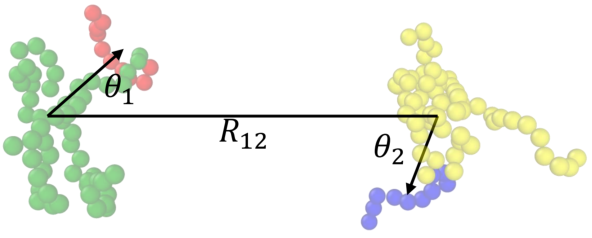}
\centering
  \caption{Schematic representation of two controlled variables: $R_{12}$, that is, distance between two centers-of-mass of the polymers, and $\theta_1, \theta_2$, that is, angles between the patch vectors and line joining the centers-of-mass of the two polymers.}
  \label{fig:rthetha}
\end{figure}

The starting point is an ensemble of independent dynamical trajectories generated from simulations. Each trajectory begins in an unbounded configuration and evolves under the prescribed dynamics until either binding occurs or the simulation terminates. Only those trajectories that successfully assemble are retained and referred to as reactive trajectories \cite{newton2015rotational}. This conditioning on successful assembly is central to the idea that the reactive path density is defined only over pathways that lead to the target state, ensuring that the analysis highlights configurations that are relevant for assembly. Each reactive trajectory is truncated at the first binding event, so that only the dynamical evolution from the unbound state to the bound state is considered. This removes post-binding fluctuations that are irrelevant to the assembly mechanism. The high-dimensional microscopic state of the system at each time step is then projected onto a low-dimensional set of collective variables (CVs). In this work, the CVs are chosen as the inter-polymer center-of-mass distance ($R_{12}$) and an angular variable $\Phi = \theta_1 + \theta_2$, where each $\theta_i$ measures the alignment between a polymer’s patch orientation and the inter-polymer separation vector (see Figure \ref{fig:rthetha}). This choice captures the essential coupling between translation and rotation during assembly\cite{newton2015rotational}. To compute the reactive path density, the CV space is discretized into bins. For each reactive trajectory, one records which bins are visited at least once before binding. Importantly, a first-visit convention is employed: if a trajectory revisits the same bin multiple times, it is counted only once for that bin. This avoids over-weighting slow or diffusive trajectories and ensures that the resulting density reflects pathway topology rather than residence times\cite{newton2015rotational}. After processing all reactive trajectories, the count in each bin is normalized by the total number of reactive trajectories, yielding the reactive path density, $n_r(R_{12}, \theta_1 + \theta_2)$. The resulting density does not integrate to unity and should not be interpreted as a probability distribution. Instead, it quantifies how important each region of CV space is for successful assembly\cite{newton2015rotational}. Regions with high reactive path density are those that most successful trajectories must pass through, whereas regions with low or zero density correspond to configurations that are dynamically accessible but irrelevant to binding.

\begin{figure}[t]
\centering
\includegraphics[scale=1]{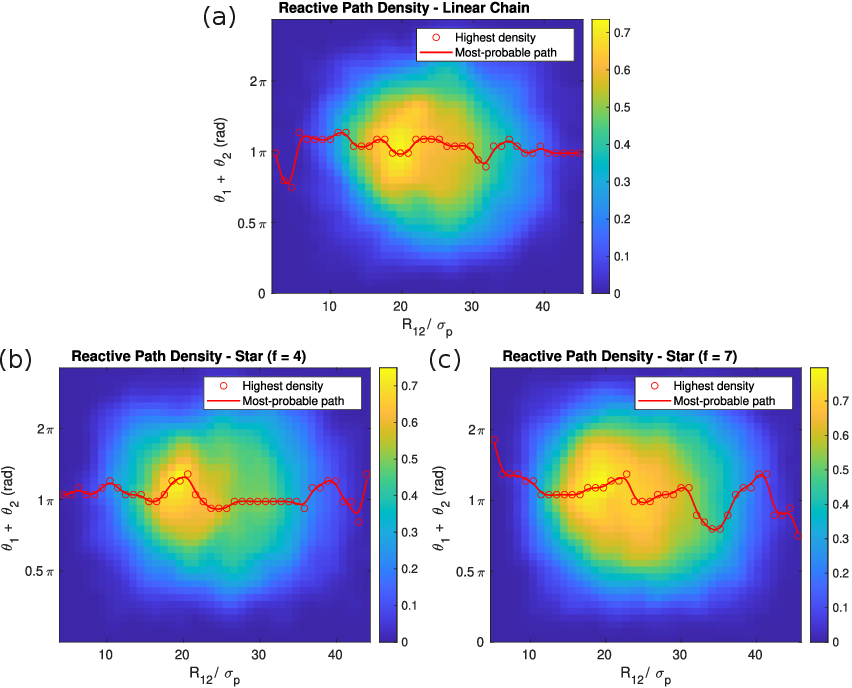}
\caption{Reactive path density: (a) linear chain, (b) star polymer with functionality four, and (c) star polymer with functionality seven.}
\label{fig:path_density}
\end{figure}

Dynamical pathways are revealed by the structure of the reactive path density. Typically, the density forms a broad, tube-like channel in CV space, reflecting the ensemble of noisy, diffusive trajectories that lead to assembly. To extract a concise representation of this channel, one identifies its ridge: for each value of $R_{12}$, the value of $\theta_1 + \theta_2$ at which the reactive path density is maximal is determined. This produces a discrete set of ridge points, which are then smoothed to obtain a continuous curve. This smooth red curve (see Figure \ref{fig:path_density}) represents the most-probable dynamical pathway, or diffusive path, followed by the system during successful assembly. Crucially, this pathway is not a single deterministic trajectory. Rather, it is a statistical summary of the ensemble of reactive trajectories, indicating the typical sequence of configurational changes conditioned on assembly\cite{newton2015rotational}. Figure \ref{fig:path_density} shows the reactive path density for the assembly of two polymers as a function of the inter-polymer separation $R_{12}$ and the combined orientational variable $\theta_1 + \theta_2$ for three different polymer architectures: (a) linear chains, (b) four-arm star polymers, and (c) seven-arm star polymers. The color map represents the reactive path density, which measures how frequently successful assembly trajectories visit a given region of collective-variable space prior to binding. Red circles indicate the ridge points corresponding to the maximum density at fixed separation, while the red curve shows the smoothed most-probable (diffusive) assembly pathway. 

Using colloidal patchy particles as a model system, \citet{newton2015rotational} demonstrated that varying the relative importance of rotational and translational diffusion profoundly alters the dynamical pathways of self-assembly. To quantify this effect, they introduced the rotational diffusion ratio as follows,
\begin{equation}
    f_{SE} = \sqrt{\frac{D_r \sigma^2}{3 D_t}}
\end{equation}
where $D_r$ and $D_t$ are the rotational and translational diffusion coefficients of the particles, respectively, and $\sigma$ denotes the particle radius. The suffix "SE" stands for Stokes-Einstein\cite{newton2015rotational}. This dimensionless ratio captures the balance between rotational and translational motion during assembly. Treating the hydrodynamic radius ($R_h$) as the effective particle size, we compute the rotational diffusion ratio as follows,
\begin{equation}
    f_{SE} = \sqrt{\frac{D_r R_h^2}{3 D_t}}
\end{equation}
The resulting values of $f_{SE}$ are listed in the final column of Table \ref{tbl:ratioSE}. \citet{newton2015rotational} showed that when the rotational diffusion ratio is small ($f_{SE} \sim 0.1$), corresponding to slow rotational diffusion, patchy particles tend to assemble along nearly straight dynamical pathways dominated by translational motion. In contrast, for large values of the ratio ($f_{SE} \sim 10$), where rotational diffusion is fast, assembly proceeds along more curved pathways in collective-variable space, reflecting frequent reorientation during approach.

For the polymer systems studied here, the calculated values of $f_{SE}$ lie in the intermediate range $ 1 \le f_{SE} \le 3$. Consequently, the observed dynamical pathways neither strictly follow a straight radial trajectory nor exhibit strongly curved behavior. Instead, they fluctuate around an approximately straight path with moderate deviations arising from orientational dynamics. Notably, star polymers with functionality $f = 7$ display more pronounced curvature in their dynamical pathways compared to polymers with lower functionality, consistent with their relatively higher rotational diffusion ratio. This observation further underscores the role of polymer architecture in tuning the balance between rotational and translational motion during self-assembly.

\section{Concluding Remarks}
In this work, we investigated the self-assembly of two patchy macromolecules modeled as linear chains and star polymers with functionalities $f=4$ and $f=7$ in an MPCD solvent. The molecular weights of the polymers were selected such that the resulting hydrodynamic radii are nearly identical, thereby matching their translational diffusion coefficients and enabling the influence of rotational diffusion on self-assembly kinetics to be isolated. The binding probability is found to depend strongly on the internal architecture of the macromolecules. Among the systems considered, linear polymers exhibit the highest binding probability, followed by four-arm and seven-arm star polymers. This trend correlates with the decreasing radius of gyration of the star polymers with an increase in functionality at constant hydrodynamic radius, indicating that macromolecular architecture influences the accessibility of the reactive patches even when the effective hydrodynamic size is nearly unchanged.

The dynamical pathways of self-assembly were characterized using reactive path density analysis and correlated with the rotational diffusion ratio, $f_{\mathrm{SE}}$. The values of $f_{\mathrm{SE}}$ lie in the intermediate regime for all three polymer architectures, indicating that the dominant reactive trajectories are neither perfectly straight nor highly curved. Among them, the seven-arm star polymer follows the most curved reactive pathways owing to its relatively higher rotational diffusivity.

Overall, we show that the internal architecture of macromolecules plays a crucial role in determining self-assembly kinetics. By isolating the effect of rotational diffusion while maintaining nearly identical translational diffusion, this study establishes a direct connection between the equilibrium rotational dynamics of individual macromolecules and their stochastic self-assembly pathways, providing molecular-level insights that may aid the rational design of self-assembling polymeric systems.

\begin{acknowledgement}

G.T. gratefully acknowledges the support received from the National Supercomputing Mission of the Department of Science and Technology (DST), India, for runtime on the PARAM Pravega high-performance computing system housed in the Supercomputing Education and Research Center-Indian Institute of Science, Bengaluru.

\end{acknowledgement}

% \begin{suppinfo}

% The following files are available free of charge.
% \begin{itemize}
% \item Supplemental Material: (S1) Diffusion of monomers, (S2) Comparison with Zimm theory, (S3) Translational diffusion at higher hydrodynamic radius, (S4) Gyration tensor, (S5) Rigid body rotation approximation, (S6) Comparison with Perrin’s formula, (S7) Role of flexibility on reorientation of polymers, (S8) Translational diffusion at constant molecular weight, (S9) Rotational diffusion at constant molecular weight, (S10) Radius of gyration vs. molecular weight, (S11) Geometrical shrinking factor vs. molecular weight, (S12) Relative shape anisotropy vs. molecular weight, (S13) Monomer density profile and intramolecular structure factor, (S14) Finite size effects on translational diffusion, and (S15) Finite size effects on rotational diffusion
% \end{itemize}

% \end{suppinfo}

%\begin{suppinfo}

%\end{suppinfo}


\bibliography{achemso-demo}

\end{document}